\documentclass[10pt, a4paper, twocolumn, aps, pra, showpacs, longbibliography, superscriptaddress]{revtex4-1}
\pdfoutput=1
\usepackage[utf8x]{inputenc}
\usepackage{ucs}
\usepackage{amsmath}
\usepackage{amsfonts}
\usepackage{amssymb}
\usepackage{makeidx}
\usepackage{cellspace,booktabs}
\usepackage{natbib}
\usepackage{lipsum}
\usepackage{bm}
\usepackage{bbm}
\usepackage{hyperref}
\hypersetup{
    colorlinks=true,
    linkcolor=blue,
    filecolor=magenta,      
    urlcolor=cyan,
}

\usepackage{color}
\definecolor{light-gray}{gray}{0.55}

\usepackage{microtype}

\usepackage{graphicx}
\usepackage{siunitx}

\begin{document}

\begin{abstract}
Thin nanomaterials are key constituents of modern quantum technologies and materials research.
Identifying specimens of these materials with properties required for the development of state of the art quantum devices is usually a complex and lengthy human task. In this work we provide a neural-network driven solution that allows for accurate and efficient scanning, data-processing and sample identification of experimentally relevant two-dimensional materials. We show how to approach classification of imperfect imbalanced data sets using an iterative application of multiple noisy neural networks. We embed the trained classifier into a comprehensive solution for end-to-end automatized data processing and sample identification.

\end{abstract}

\date{\today}
\author{Eliska Greplova}
\affiliation{Institute for Theoretical Physics, ETH Zurich, CH-8093, Switzerland}
\author{Carolin Gold}
\thanks{These two authors contributed equally}
\affiliation{Solid State Physics Laboratory, ETH Zurich, CH-8093 Zurich, Switzerland}
\author{Benedikt Kratochwil}
\thanks{These two authors contributed equally}
\affiliation{Solid State Physics Laboratory, ETH Zurich, CH-8093 Zurich, Switzerland}
\author{Tim Davatz}
\affiliation{Solid State Physics Laboratory, ETH Zurich, CH-8093 Zurich, Switzerland}
\author{Riccardo Pisoni}
\affiliation{Solid State Physics Laboratory, ETH Zurich, CH-8093 Zurich, Switzerland}
\author{Annika Kurzmann}
\affiliation{Solid State Physics Laboratory, ETH Zurich, CH-8093 Zurich, Switzerland}
\author{Peter Rickhaus}
\affiliation{Solid State Physics Laboratory, ETH Zurich, CH-8093 Zurich, Switzerland}
\author{Mark H. Fischer}
\affiliation{Institute for Theoretical Physics, ETH Zurich, CH-8093, Switzerland}
\affiliation{Department of Physics, University of Zurich, Winterthurerstrasse 190, 8057 Zurich, Switzerland}
\author{Thomas Ihn}
\affiliation{Solid State Physics Laboratory, ETH Zurich, CH-8093 Zurich, Switzerland}
\author{Sebastian Huber}
\affiliation{Institute for Theoretical Physics, ETH Zurich, CH-8093, Switzerland}

\title{Fully automated identification of 2D material samples}

\maketitle

\section{INTRODUCTION}

Since the isolation of graphene \cite{Novoselov2004}, two-dimensional (2D) materials constitute an active area of research with numerous applications in optoelectronics ~\cite{Xia2014,Bonaccorso2010,Koppens2014,Wang2012} and they serve as basic building blocks for a wide range of quantum devices \cite{Novoselov2016a}. This is owed to the van der Waals stacking technique~\cite{Dean2010,Geim2013,frisenda_recent_2018} which allows for drastic modifications of the band structure by stacking different materials, varying the number of layers or the twist between the layers~\cite{Cao2018,Yankowitz2019,Conley2013}.

High-quality Van-der-Waals devices consist of flakes that are typically prepared by mechanical exfoliation \cite{Novoselov2004}. Suitable flakes are identified via visual inspection in an optical microscope \cite{Blake2007}. The shape,
size and homogeneity determine whether the flake is suitable
for further processing. However, the observed difference in contrast and color of a flake with respect to the background does not only depend on its thickness and material, but also on the substrate that is used and on the settings of the microscope.
This large parameter space makes the identification of usable flakes tedious and, while there exist proposed algorithmic solutions~\cite{Li2019, Masubuchi2019,Masubuchi2018auto, Funke2017,gorbachev2011hunting,li2013rapid,lin2018intelligent,saito2019deep}, a sufficiently general and fast algorithm is difficult to formulate.

In the present work we address the issue of fast and reliable identification of 2D material samples using a supervised machine learning algorithm. Machine learning has proven to be a successful method for addressing the classification of the large noisy data sets in many areas both in science and engineering~\cite{Kotsiantis2007, Mehta2019, Zoph2018, Hu2018,Nielsen2015}. In this context, the problem of 2D sample identification is particularly daunting due to the lack of data (suitable flakes are in general rare and hard to find). Moreover, the challenge we are addressing here is not only that of classification alone but a search for an end-to-end algorithm that enables scanning of the samples, pre-processing, fast classification and identification of the suitable flakes in the frame of reference of the original sample.

Dealing with realistic data sets can overwhelm even established machine learning methods, since the collected data can be insufficient, unbalanced, mislabeled and have high amount of noise. Here, we discuss the particular problem of the efficient collection of suitable hexagonal Boron Nitride (hBN) flakes. This problem is an example of the broader class of noisy classification problems: hBN is a crucial element of quantum devices experiments, but its collection is difficult, as appropriate flakes are exceedingly rare and qualitatively very different from each other. Finding hBN flakes typically requires many hours of expert human labor or advanced image processing software. The ideal flake is uniform, isolated from other flakes and has a certain thickness that is suited for a specific experiment. hBN is typically used as a flat substrate (in this case the thickness is not relevant) or as a gate dielectric (with thickness between $10$ and $90$ nm). In addition, different users may prefer slightly different flakes.
These are issues that make machine learning and particularly neural networks highly apt for this task: the goal of machine learning is to extract general features from a possibly limited set of data. An additional advantage of machine learning methods is that, once trained, the model can be applied to new data in a matter of seconds as opposed to the repeated run of an computationally heavy algorithm.

In this paper we present a fully automated solution for identifying suitable hBN flakes on a wafer. We automatize a setup consisting of a microscope, a camera and a sample stage in a glovebox to scan wafers carrying exfoliated hBN flakes. In a pre-processing step flakes on the wafer are detected and uniformly formatted. A multi-layer convolutional neural-network is employed to identify promising flakes. In a last step the algorithm captures pictures of the promising flakes in a higher resolution. We report a success rate of our algorithm which is comparable to that of a human operator, while being significantly faster.

\begin{figure}
\includegraphics[scale=0.42]{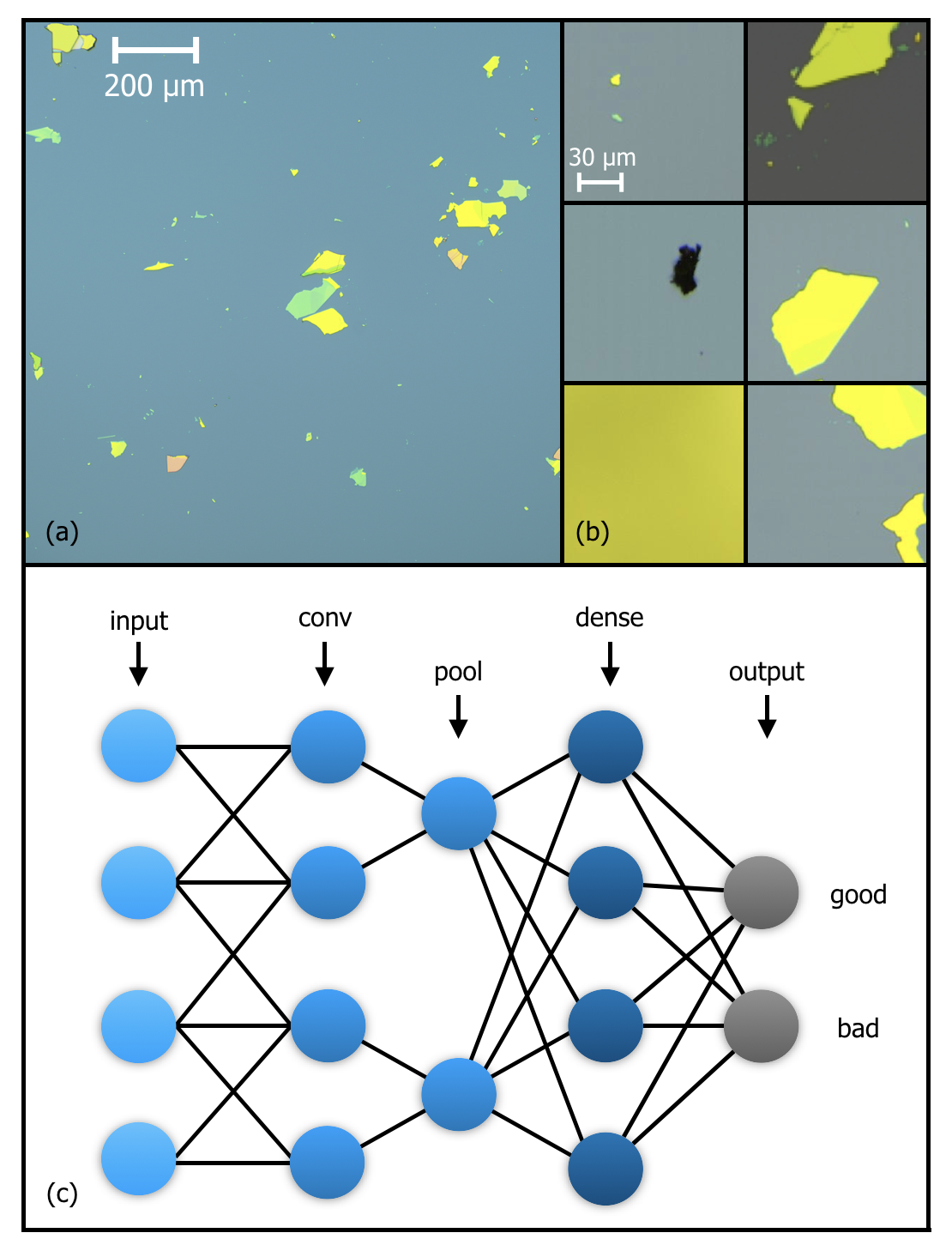}
\caption{Collection and processing of data: (a) typical microscope image of hBN, (b) examples of the pre-processed images, first column represents examples picked up by our pre-processing algorithm that are not hBN flakes, the second column constitutes of examples picked-up by pre-processing constituting useful hBN flakes, (c) architecture of the used neural network.}\label{fig:samples_network}
\end{figure}

\section{Algorithm}
Deep neural networks are the state of art candidates for solving diverse classification problems \cite{Dreiseitl2002, Wan1990neural,Krizhevsky2012imagenet,Cirecsan2012multi,Carrasquilla2017machine,Greplova2017}. Here we want to utilize a neural network to classify data on the fly, while they are being collected by the microscope and return the coordinates on the experimentally relevant parts of the sample as fast as possible. In this section, we show how to automatize this process, such that a human expert does not have to identify the suitable samples manually, but only needs to proceed, when the suitable samples coordinates have been automatically identified for them.

Our approach to collecting and classifying data consists of the following steps: (a) scanning, or automated data collection, (b) manual labeling of the obtained data, (c) pre-processing the data, (d) training of the machine learning model, (e) applying the trained neural network to the new data, (f) identifying the coordinates of the flakes of interest on the sample. Steps (b) and (d) are only performed once and skipped when the trained model is being applied to the new raw data.

\emph{Scanning.} Step (a) is accomplished by an automated setup consisting of an off-the-shelf microscope, camera and sample stage in a glovebox. This setup scans wafers of exfoliated hBN and captures an image at fixed positions. Prior to the scan, the scanning area and focal plane of the sample are defined through user input. This step takes minutes and, crucially, is independent of wafer size. At each position the camera automatically adjusts the focus according to the user input and captures an image. The machine learning algorithm we explain below processes these images in parallel to the scan, thus avoiding any downtime of the setup. At a microscope magnification of $10 \times$ this procedure is capable of scanning $190$s/cm$^2$, significantly faster than the average human operator. Furthermore, due to the automation, the use of human and machine time can be significantly optimized. An example of a typical image obtained by sample scanning is shown in Fig.~\ref{fig:samples_network}(a).

\emph{Labeling.} For training of the machine learning algorithm, the images obtained in step (a) are then labeled during step (b) by human operators who identify suitable flakes. The fact that different people label the data gives rise to a larger variability in the labeling of the data. The labeling and training step is only performed once. To facilitate the labeling process we developed a graphical interface, where the user simply clicks on the samples they consider suitable for further processing. The coordinates of the flakes within the given picture as well as within the whole wafer is saved for future identification.

\emph{Pre-processing.} Having collected (and, in the case of the training phase of the algorithm, labeled) data, the pre-processing step (c) is applied. The minimal pre-processing algorithm identifies potential candidates for the usable flakes and formats them uniformly for the classification step. We use the Python Image Library (PIL) to check for large standard deviation in the data. Whenever the standard deviation exceeds a chosen threshold, the algorithm cuts out a fixed-size square around the given point. The square images ($80\times80$ pixels) then become elements of the training set for the neural network. Already at this stage, every flake is uniquely identified by its coordinates with respect to the coordinate system spanned by three corners for the wafer. This identifier is kept throughout all further steps. Examples of suitable flakes are shown in the right hand column of Fig. \ref{fig:samples_network}(b), while examples of unsuitable flakes are in the left hand column.

\begin{figure*}
\includegraphics[scale=0.9]{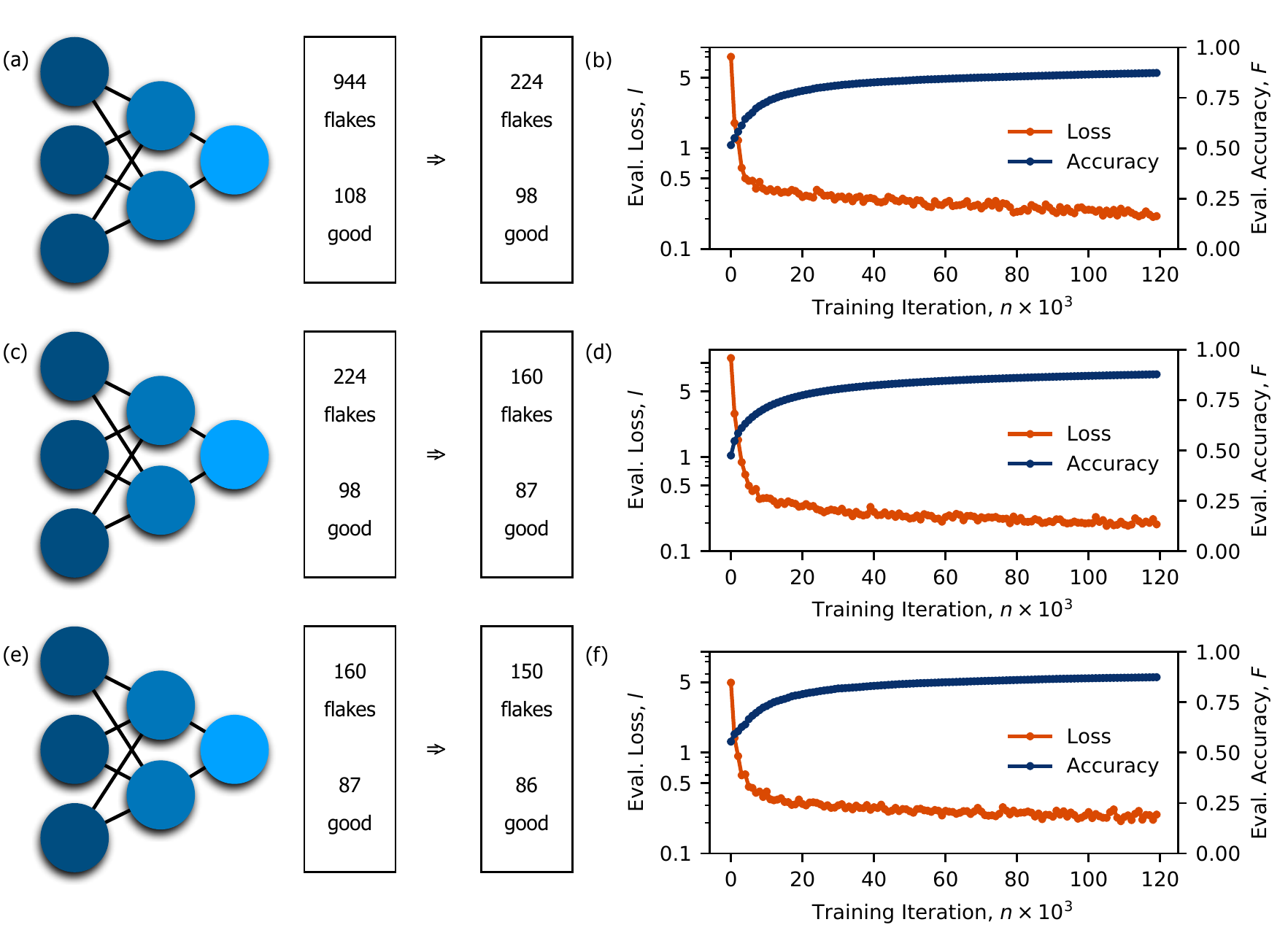}
\caption{Iterative application of the multiple networks: panels (a), (c), (e) show how the iterative procedure filters out bad flakes on the concrete example of a small set constituting of 944 flakes. In each step we display the total number of found flakes and total number of flakes labeled as 'good'. Panels (b), (d), (f) show validation loss and accuracy of each of the networks in (a), (c), (e) respectively as a function of the training step.}
\label{fig:big_fig}
\end{figure*}

We use the pre-processing part of the algorithm to assign two labels to the training images. When the square image created around an interesting point in the data contains a suitable flake (identified in step (b)), that image is labeled as `good'. Whenever the given square image does not contain any suitable flakes, it is labeled as `bad'.

The data set contains a large number of unsuitable flakes that are not labeled as `bad' by the standard deviation filter. The reason is that the used criterion has to be sufficiently loose and flexible such that no good flake would be missed independent of its size, thickness or the color of the wafer. This requirement is met by choosing an appropriate threshold value the standard deviation filtering. 

We artificially increase the number of `good' flakes in the training set by using image rotations and mirror reflections, such that each `good' flake enters the training set in 6 different variations. This choice significantly shortens the preparation of the training set that contains a sufficient amount of 'good' flakes for successful training of our model.

As seen in Fig. \ref{fig:samples_network}(b), the examples of both  `good' and `bad' flakes are very diverse. On one hand, flakes that are too small, broken or simply just dust or pictures taken outside of the wafer all have to be identified as `bad'. On the other hand the flakes of various thicknesses and shapes have to be recognized as good even against the background whose color can vary depending on the wafer. We are interested in flakes that have a thickness between 10 and 90 nm, resulting in blue, green and yellow flakes. This diversity of the different flakes has immediate consequences for the construction of the classification mechanism. 

\emph{Training.} In step (c), the set of training set images was created. The training step (d) builds a reliable binary classification model able to distinguish two classes of flakes: `good'  or `bad'. Images with the label `good' contain a flake that is a potential candidate for further processing steps. The label `bad' is for all other images of flakes. In the following we provide a coarse description of the algorithm.

The architecture of the network in use is shown in Fig. \ref{fig:samples_network}(c).  We employ a deep network consisting of 4 convolutional layers and one dense layer. The convolutional layers have $64$, $64$, $128$, $256$ filters respectively. The dense layer has $256$ neurons (for further details see the Appendix \ref{app:net}). The output of the network is the probability distribution between two classes `good' and `bad' in the output layer. In order to obtain the weights for the different neurons in the network it was trained using the training set.

After pre-processing step (c), the data set is highly imbalanced, meaning that the number of flakes with the label `good' is very low compared to the number of flakes with the label `bad'. Picking such an imbalanced dataset for training would result in a network that labels all the flakes with the label `bad', not learning any feature of the flakes with the label `good'. On the other hand, downsampling the number of `bad' flakes would lead to a loss of the variability of the dataset. 

We overcome these issues by forming balanced sets, the so-called batches, out of the labeled data and by creating an iterative protocol consisting of multiple neural networks. In particular, in each training step the network is fed a batch that consists of randomly selected $50\%$ `good' and $50\%$ `bad' flakes. Since the number of `good' flakes is small, one has to be careful not to over-fit the features of the good flakes. We overcome this constraint by stopping the training while the network is still relatively noisy and has an accuracy around $90\%$. At this point we train further networks separately on the same training data. In our case we find that applying $3$ separate models is optimal. When generalizing to other materials this number may differ depending on the particular training set: rarer or denser distribution of the desirable samples or different pre-processing criteria will influence it.
If a given flake consecutively passes the $3$ separately trained noisy classifiers there is a high probability that it is worth to be inspected by a human expert. The scheme of the iterative model is illustrated in Fig.~\ref{fig:big_fig}.

The iterative application of consecutive models proves successful in avoiding misclassified flakes. We show below that while a single neural net would misclassify a relatively large number of `bad' flakes as `good' simply because of their statistical significance, it is increasingly unlikely that multiple models will misclassify the \emph{same} `bad' flakes as `good'. 

\emph{Application of the model.} Once the model is trained, we can apply it to any pre-processed data set that was not part of the original training data. The unique coordinates of each flake identified as `good' by the algorithm allows us to easily navigate to any flake under the microscope and to use it for further processing or device fabrication.

\section{Results}

Each of the three networks was trained on approximately $10^6$ flakes. Of these, about $10^4$ are labeled as good. We use, for the training, the batches of $200$ flakes at a time. For each of these batches we run a back-propagation algorithm in order to adjust the weights connecting the neurons in the network. We repeat this process $1.2\times10^5$ times.  These are the so-called training steps. We evaluate the performance of the model every $1000$ training steps on a fixed evaluation batch consisting of $10\%$ of the total number of `good' flakes and an equal number of `bad' flakes.

The evaluation loss and accuracy of all models are shown in Fig.~\ref{fig:big_fig} (b), (d), (f). The evaluation loss is a cross-entropy between the label ('good' or 'bad') network determined and the label determined by an expert. The accuracy is the percentage of correctly identified flakes within the chosen evaluation batch (the evaluation batch is different for the different neural networks). The accuracy reached at the end of the training is between $87\%-88\%$ as shown in Fig.~\ref{fig:big_fig} (b), (d), (f). While the loss for the trained models keeps decreasing towards zero, the training is purposefully finished at its nonzero finite value as explained in Sec. II (see also Appendix \ref{app:net}).
\begin{figure}
\includegraphics[scale=0.5]{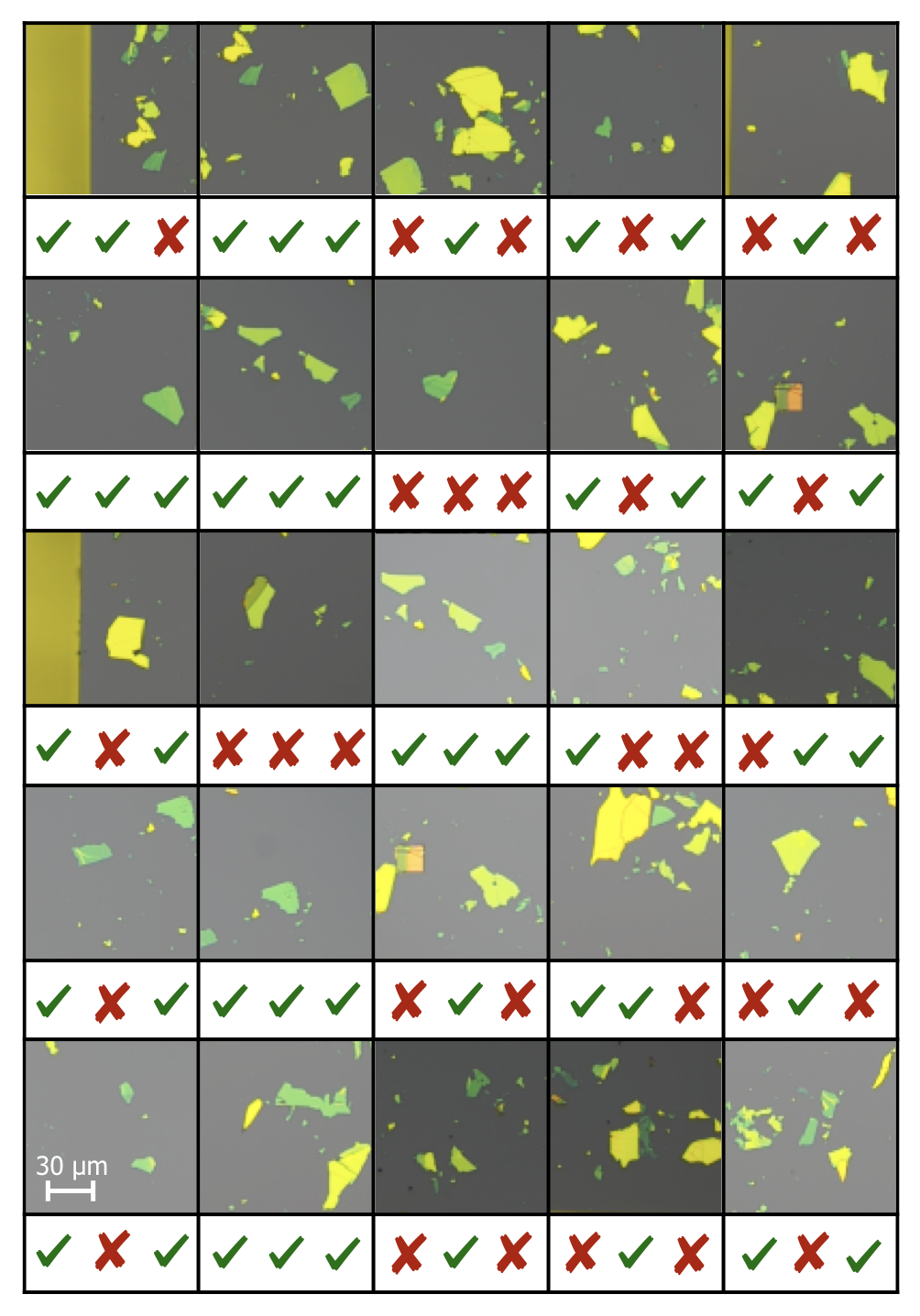}
\caption{Validation by human operator: First 25 flakes of the set for human validation are shown with evaluation by three different operators: 23 flakes are denoted to be suitable by at least one operator, 2 flakes are dismissed by all three operators.}
\label{fig:human}
\end{figure}

The performance of the protocol is shown in Fig.~\ref{fig:big_fig}. In the panel (a), the first network is applied on pre-processed un-labeled data ($944$ flakes in our case out of which $108$ are `good'. The network does not have access to this information). After the first application of the network we obtain $224$ flakes out of which $98$ are `good' as determined by a human expert. We take these flakes and use them as input for the second network and so on. After the application of the third network we obtain $150$ flakes out of which $86$ are labeled as `good' by an expert. The application of the networks takes a few seconds and it puts the human operator into a situation where approximately $55\%$ of the flakes will be suitable for further experimental use (as opposed to less than $1\%$ that would be recoverable from the direct data processing).

To evaluate the success of the protocol on the new unlabeled data we asked $3$ human operators to independently evaluate a set of $100$ flakes that the network selected as `good'. The human operators denoted $73$, $51$, $45$ flakes as `good'. However, only $7$ of the flakes have been denoted as `bad' by all three experts. This is consistent with the benchmarks we obtained from the validation procedure of our protocol that led us to expect about $55\%$ of the flakes would be of further use. The human validation is illustrated in Fig.~\ref{fig:human}. One can observe considerable variability in the human assessment of the flakes and very small amount of images that have been considered mislabeled by all three operators.

\begin{figure}
\includegraphics[scale=0.45]{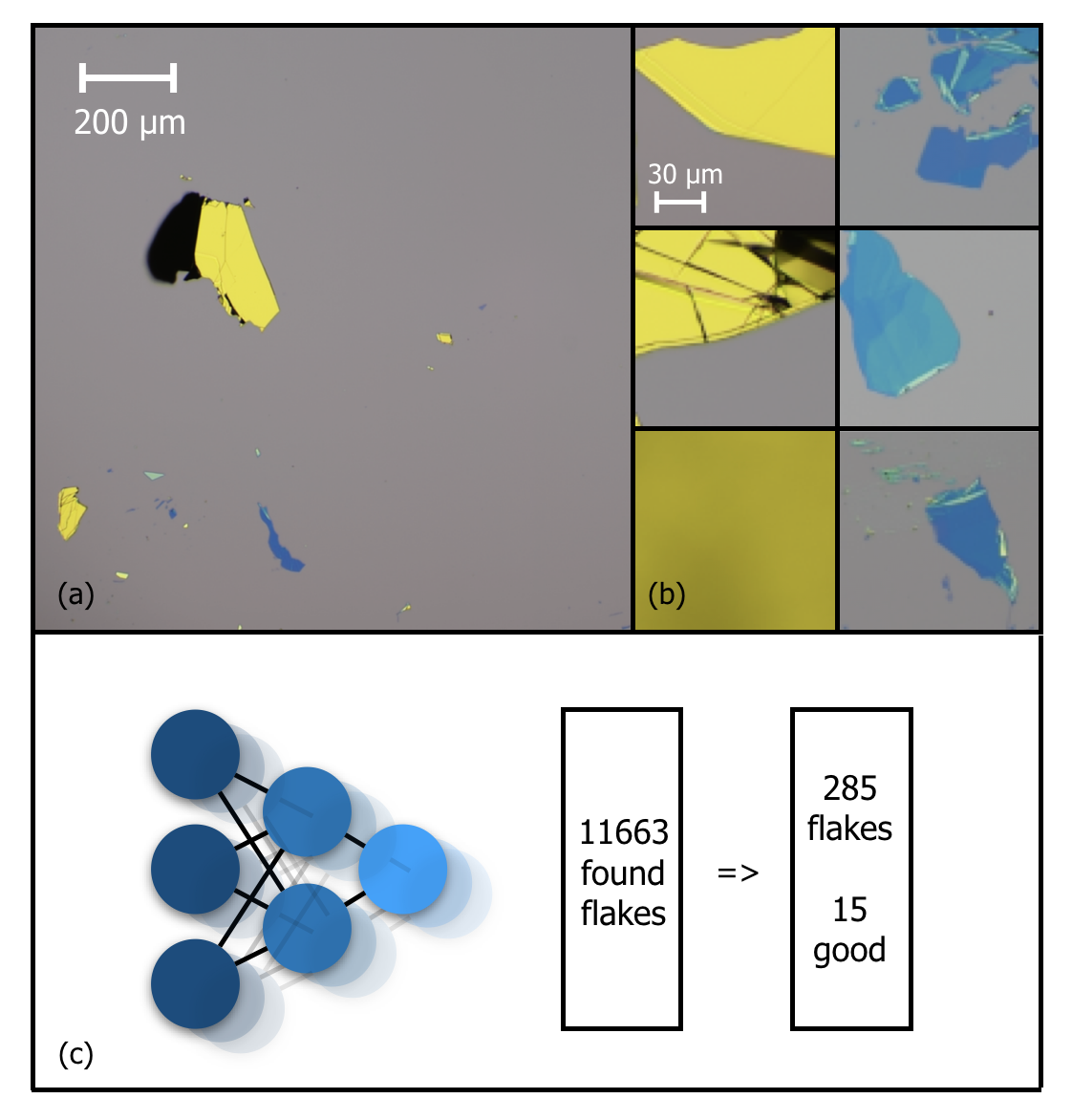}
\caption{Testing the performance on the network on different material: panel (a) shows example of picture of graphite wafer, panel (b) shows examples of bad (left) and good (right) graphite flakes, panel (c) shows results of application of our protocol on $11663$ flakes obtained by scanning single graphite wafer.}
\label{fig:graphite}
\end{figure}

Neural networks are known for their ability to generalize knowledge. While our algorithm performs well on hBN it was trained on, we can ask whether we can utilize the model we trained also for another material. We applied the model already trained on hBN on graphite. Graphite plays a similarly important role for quantum materials experiments as hBN. In Fig.~\ref{fig:graphite}, we show an  example of the typical microscope picture of a graphite sample (panel (a)), examples of `bad' and `good' flakes (panel (b)) and performance of the network on test set of approximately $11 000$ flakes (panel (c)). The network outputs $285$ flakes out of which $15$ (which corresponds to approximately $6\%$) were selected as `good' by two independent human operators. The success here is considerably lower compared to $55\%$ success rate in case of hBN. The reason is that only a specific thickness range of graphite is of experimental interest and the network selects quite broad range of thicknesses. We nevertheless succeeded in reducing significantly the task of the human operator. In addition to that, this results suggest that the network could be straightforwardly partially retrained for another material.

\section{Discussion}
We presented a fully automated method to evaluate the quality and suitability of 2D-material specimens for sample fabrication. Including every step from data-collection and pre-processing to the evaluation of suitable flakes, the algorithm removes 99\% of redundant data without human assistance. From the remaining 1\% of the flakes labeled as `good' by the algorithm, approximately $55\%$ are actually suitable for further experimental use. The exact percentage of the latter can vary significantly depending on the needs of the respective application. In our case, experiments with twisted bilayer graphene \cite {rickhaus2018transport} require thick (\SI{90}{nm}) and huge flakes ($\SI{100}{\micro m}$), while samples used for nano devices \cite{eich2018spin,eich2018coupled,overweg2017electrostatically} require thin (about \SI{20}{nm}) and small ($\SI{20}{\micro m}$) flakes. Hence, almost all flakes, identified by the network are usable for one of our experiments, as seen in the human validation in Fig.~\ref{fig:human}.

We trained and successfully tested our algorithm on the detection of hBN flakes. We applied the model without retraining on graphite samples. While the accuracy is significantly lower than for hBN, the algorithm is still successfully eliminating a high percentage of unsuitable flakes and leaves a human operator with the order of hundred flake candidates, out of which $5\%$ are suitable for further processing. Given different thickness of graphite and therefore requirements on the color of the flakes, it is not surprising that the model does not show as high a success rate as for the material it was trained on. It however provides an indication that our model could be partially retrained on graphene for the required thickness range. Extending algorithmic strategies presented here to other 2D materials promises to facilitate the fabrication of a large variety of different 2D-material based devices \cite{ferrari2015science,jariwala2014emerging,fiori2014electronics,liu2016van}. The ready-to-use algorithm is available at \href{http://github.com/cmt-qo/cm-flakes}{http://github.com/cmt-qo/cm-flakes}. We provide both API for the automation of the data collection as well as pre-trained models that can be applied on newly collected raw data.

\section*{Acknowledgements}
We gratefully acknowledge financial support from the Swiss National Science Foundation, the NCCR QSIT. This work has received funding from the European Research Council under grant agreement no. 771503.
We acknowledge helpful discussions with Klaus Ensslin and Yuya Shimazaki.
We acknowledge Rebekka Garreis, Chuyao Tong and Yongjin Lee for contributing to the labeled database. We thank Peter Blake for his help with the data acquisition from the microscope.

\appendix

\section{Neural Net Architecture}
\label{app:net}

Neural networks are nonlinear functions of many parameters (weights and biases of each neuron) mapping input to output such that a certain cost function between input and output is minimized. This minimization is achieved by optimizing the weights and biases of the neuron using the back-propagation method \cite{Rumelhart1988}.
In this work we use $3$ neural networks separately trained on the same highly imbalanced training set consisting of approximately $10^6$ samples. Roughly $10\%$ of these samples correspond to flakes of 2D material that is labeled as 'good', or is of further experimental use.
Not all these approximately $10^5$ flakes are original, most of them have been created by mirroring and rotations of the rare experimentally collected good flakes.
We train on batches of $200$ flakes ($100$ of them 'good', $100$ of them 'bad'). As discussed in the main text, this balance is crucial for optimizing the network variables correctly. We train the network on the training set consisting of  $1.2\times10^5$ samples (flakes). We evaluate the model every $1000$ training steps on a fixed evaluation batch consisting on $10\%$ of the total number of `good' flakes and equal number of `bad' flakes. The accuracy at the end of the training is between $87\%-88\%$ as shown in Fig.~\ref{fig:big_fig} (b), (d), (f). The cost function we aim to minimize by training is the cross-entropy between the label ('good' or 'bad') determined by the network ($y^{\textit{output}}$) and the true label assigned by an experimentalist ($y^{\textit{target}}$) can be expressed as
\begin{equation}
H(y^{\textit{target}},y^{\textit{output}})=-\sum_j y^{\textit{target}}\ln y^{\textit{output}},
\label{eq:cross_entropy}
\end{equation}
where $y^{\textit{target}},y^{\textit{output}}\in\{\textit{good},\textit{bad}\}$. By trying to minimize Eq.~\eqref{eq:cross_entropy} we penalize the misclassification of both 'good' and 'bad' flakes and force the network to find the weights that leads to recovery of the correct classification label.

\begin{table}[]
\begin{tabular}{@{}lllll@{}}
\toprule
 type of layer & filters &  kernel size & strides & neurons  \\ \midrule
 conv& 64 &  5& 2 & NA \\
 conv& 64 & 3 & 1 & NA \\
conv & 128 & 3 &  1&NA \\
conv & 256 & 3 &  1&NA \\
dense & NA & NA & NA & 256 \\ \bottomrule
\end{tabular}
\caption{Summary of hyperparameters of the neural net.}\label{tab:hyper}
\end{table}

Let us describe in more detail the structure of our models. As mentioned above each of the networks we use consists of $4$ convolutional layers and $1$ dense layer. The hyperparameters for the layers we use are summarized in Table \ref{tab:hyper}. The first convolution layer is defined with a larger kernel and stride and a low amount of filters. The deeper layers contain a larger number of filters and smaller kernels. The dense layer contains $256$ neurons. These neurons are then connected to two output neurons that contain a probability that the given flake is good or bad respectively. The activation function for all the layers except the last is a rectified linear unit (ReLU), while  the last layer contains a softmax activation function that transforms the values of the neurons into the probablity distribution. For a detailed overview of both the theoretical aspects and the practical implementation of neural networks we refer the reader to \cite{Geron2017,Nielsen2015}.

\bibliographystyle{unsrt}
\bibliography{flakes-bibliography}

\begin{thebibliography}{10}

\bibitem{Novoselov2004}
KS~Novoselov, A~Mishchenko, A~Carvalho, and AH~Castro Neto.
\newblock 2d materials and van der waals heterostructures.
\newblock {\em Science}, 353(6298):aac9439, 2016.

\bibitem{Xia2014}
Fengnian Xia, Han Wang, Di~Xiao, Madan Dubey, and Ashwin Ramasubramaniam.
\newblock {Two-dimensional material nanophotonics}.
\newblock {\em Nat. Photonics}, 8:899, nov 2014.

\bibitem{Bonaccorso2010}
F~Bonaccorso, Z~Sun, T~Hasan, and A~C Ferrari.
\newblock {Graphene photonics and optoelectronics}.
\newblock {\em Nat. Photonics}, 4:611, aug 2010.

\bibitem{Koppens2014}
F~H~L Koppens, T~Mueller, Ph. Avouris, A~C Ferrari, M~S Vitiello, and M~Polini.
\newblock {Photodetectors based on graphene, other two-dimensional materials
  and hybrid systems}.
\newblock {\em Nat. Nanotechnol.}, 9:780, oct 2014.

\bibitem{Wang2012}
Qing~Hua Wang, Kourosh Kalantar-Zadeh, Andras Kis, Jonathan~N Coleman, and
  Michael~S Strano.
\newblock {Electronics and optoelectronics of two-dimensional transition metal
  dichalcogenides}.
\newblock {\em Nat. Nanotechnol.}, 7:699, nov 2012.

\bibitem{Novoselov2016a}
K~S Novoselov, A~Mishchenko, A~Carvalho, A~H~Castro Neto, and Oxford Road.
\newblock {2D materials and van der Waals heterostructures}.
\newblock {\em Science (80-. ).}, 353(6298):aac9439, 2016.

\bibitem{Dean2010}
C~R Dean, A~F Young, I~Meric, C~Lee, L~Wang, S~Sorgenfrei, K~Watanabe,
  T~Taniguchi, P~Kim, K~L Shepard, and J~Hone.
\newblock {Boron nitride substrates for high-quality graphene electronics}.
\newblock {\em Nat. Nanotechnol.}, 5:722, aug 2010.

\bibitem{Geim2013}
A~K Geim and I~V Grigorieva.
\newblock {Van der Waals heterostructures}.
\newblock {\em Nature}, 499:419, jul 2013.

\bibitem{frisenda_recent_2018}
Riccardo Frisenda, Efrén Navarro-Moratalla, Patricia Gant, David Pérez~De
  Lara, Pablo Jarillo-Herrero, Roman~V. Gorbachev, and Andres
  Castellanos-Gomez.
\newblock Recent progress in the assembly of nanodevices and van der {Waals}
  heterostructures by deterministic placement of 2d materials.
\newblock {\em Chemical Society Reviews}, 47(1):53--68, January 2018.

\bibitem{Cao2018}
Yuan Cao, Valla Fatemi, Shiang Fang, Kenji Watanabe, Takashi Taniguchi,
  Efthimios Kaxiras, and Pablo Jarillo-Herrero.
\newblock Unconventional superconductivity in magic-angle graphene
  superlattices.
\newblock {\em Nature}, 556(7699):43, 2018.

\bibitem{Yankowitz2019}
Matthew Yankowitz, Shaowen Chen, Hryhoriy Polshyn, Yuxuan Zhang, K~Watanabe,
  T~Taniguchi, David Graf, Andrea~F Young, and Cory~R Dean.
\newblock Tuning superconductivity in twisted bilayer graphene.
\newblock {\em Science}, 363(6431):1059--1064, 2019.

\bibitem{Conley2013}
Hiram~J Conley, Bin Wang, Jed~I Ziegler, Richard~F Haglund~Jr, Sokrates~T
  Pantelides, and Kirill~I Bolotin.
\newblock Bandgap engineering of strained monolayer and bilayer mos2.
\newblock {\em Nano letters}, 13(8):3626--3630, 2013.

\bibitem{Blake2007}
P.~Blake, E.~W. Hill, A.~H. Castro~Neto, K.~S. Novoselov, D.~Jiang, R.~Yang,
  T.~J. Booth, and A.~K. Geim.
\newblock Making graphene visible.
\newblock {\em Applied Physics Letters}, 91(6):063124, 2007.

\bibitem{Li2019}
Yuhao Li, Yangyang Kong, Jinlin Peng, Chuanbin Yu, Zhi Li, Penghui Li, Yunya
  Liu, Cun-Fa Gao, and Rong Wu.
\newblock Rapid identification of two-dimensional materials via machine
  learning assisted optic microscopy.
\newblock {\em Journal of Materiomics}, 2019.

\bibitem{Masubuchi2019}
Satoru Masubuchi and Tomoki Machida.
\newblock Classifying optical microscope images of exfoliated graphene flakes
  by data-driven machine learning.
\newblock {\em npj 2D Materials and Applications}, 3(1):4, 2019.

\bibitem{Masubuchi2018auto}
Satoru Masubuchi, Masataka Morimoto, Sei Morikawa, Momoko Onodera, Yuta
  Asakawa, Kenji Watanabe, Takashi Taniguchi, and Tomoki Machida.
\newblock Autonomous robotic searching and assembly of two-dimensional crystals
  to build van der waals superlattices.
\newblock {\em Nature communications}, 9(1):1413, 2018.

\bibitem{Funke2017}
S~Funke, U~Wurstbauer, B~Miller, A~Matkovi{\'c}, A~Green, A~Diebold,
  C~R{\"o}ling, and PH~Thiesen.
\newblock Spectroscopic imaging ellipsometry for automated search of flakes of
  mono-and n-layers of 2d-materials.
\newblock {\em Applied Surface Science}, 421:435--439, 2017.

\bibitem{gorbachev2011hunting}
Roman~V Gorbachev, Ibtsam Riaz, Rahul~R Nair, Rashid Jalil, Liam Britnell,
  Branson~D Belle, Ernie~W Hill, Kostya~S Novoselov, Kenji Watanabe, Takashi
  Taniguchi, Andre~K Geim, and Peter Blake.
\newblock Hunting for monolayer boron nitride: optical and raman signatures.
\newblock {\em Small}, 7(4):465--468, 2011.

\bibitem{li2013rapid}
Hai Li, Jumiati Wu, Xiao Huang, Gang Lu, Jian Yang, Xin Lu, Qihua Xiong, and
  Hua Zhang.
\newblock Rapid and reliable thickness identification of two-dimensional
  nanosheets using optical microscopy.
\newblock {\em ACS nano}, 7(11):10344--10353, 2013.

\bibitem{lin2018intelligent}
Xiaoyang Lin, Zhizhong Si, Wenzhi Fu, Jianlei Yang, Side Guo, Yuan Cao, Jin
  Zhang, Xinhe Wang, Peng Liu, Kaili Jiang, and Weisheng Zhao.
\newblock Intelligent identification of two-dimensional nanostructures by
  machine-learning optical microscopy.
\newblock {\em Nano Research}, 11(12):6316--6324, 2018.

\bibitem{saito2019deep}
Yu~Saito, Kento Shin, Kei Terayama, Shaan Desai, Masaru Onga, Yuji Nakagawa,
  Yuki~M Itahashi, Yoshihiro Iwasa, Makoto Yamada, and Koji Tsuda.
\newblock Deep learning-based quality filtering of mechanically exfoliated 2d
  crystals.
\newblock {\em arXiv preprint arXiv:1907.03239}, 2019.

\bibitem{Kotsiantis2007}
Sotiris~B Kotsiantis, I~Zaharakis, and P~Pintelas.
\newblock Supervised machine learning: A review of classification techniques.
\newblock {\em Emerging artificial intelligence applications in computer
  engineering}, 160:3--24, 2007.

\bibitem{Mehta2019}
Pankaj Mehta, Marin Bukov, Ching-Hao Wang, Alexandre~GR Day, Clint Richardson,
  Charles~K Fisher, and David~J Schwab.
\newblock A high-bias, low-variance introduction to machine learning for
  physicists.
\newblock {\em Physics Reports}, 2019.

\bibitem{Zoph2018}
Barret Zoph, Vijay Vasudevan, Jonathon Shlens, and Quoc~V Le.
\newblock Learning transferable architectures for scalable image recognition.
\newblock In {\em Proceedings of the IEEE conference on computer vision and
  pattern recognition}, pages 8697--8710, 2018.

\bibitem{Hu2018}
Jie Hu, Li~Shen, and Gang Sun.
\newblock Squeeze-and-excitation networks.
\newblock In {\em Proceedings of the IEEE conference on computer vision and
  pattern recognition}, pages 7132--7141, 2018.

\bibitem{Nielsen2015}
Michael~A Nielsen.
\newblock {\em Neural networks and deep learning}, volume~25.
\newblock Determination press San Francisco, CA, USA:, 2015.

\bibitem{Dreiseitl2002}
Stephan Dreiseitl and Lucila Ohno-Machado.
\newblock Logistic regression and artificial neural network classification
  models: a methodology review.
\newblock {\em Journal of biomedical informatics}, 35(5-6):352--359, 2002.

\bibitem{Wan1990neural}
Eric~A Wan.
\newblock Neural network classification: A bayesian interpretation.
\newblock {\em IEEE Transactions on Neural Networks}, 1(4):303--305, 1990.

\bibitem{Krizhevsky2012imagenet}
Alex Krizhevsky, Ilya Sutskever, and Geoffrey~E Hinton.
\newblock Imagenet classification with deep convolutional neural networks.
\newblock In {\em Advances in neural information processing systems}, pages
  1097--1105, 2012.

\bibitem{Cirecsan2012multi}
Dan Cire{\c{s}}an, Ueli Meier, Jonathan Masci, and J{\"u}rgen Schmidhuber.
\newblock Multi-column deep neural network for traffic sign classification.
\newblock {\em Neural networks}, 32:333--338, 2012.

\bibitem{Carrasquilla2017machine}
Juan Carrasquilla and Roger~G Melko.
\newblock Machine learning phases of matter.
\newblock {\em Nature Physics}, 13(5):431, 2017.

\bibitem{Greplova2017}
Eliska Greplova, Christian~Kraglund Andersen, and Klaus M{\o}lmer.
\newblock Quantum parameter estimation with a neural network.
\newblock {\em arXiv preprint arXiv:1711.05238}, 2017.

\bibitem{rickhaus2018transport}
Peter Rickhaus, John Wallbank, Sergey Slizovskiy, Riccardo Pisoni, Hiske
  Overweg, Yongjin Lee, Marius Eich, Ming-Hao Liu, Kenji Watanabe, Takashi
  Taniguchi, Thomas Ihn, and Klaus Ensslin.
\newblock Transport through a network of topological channels in twisted
  bilayer graphene.
\newblock {\em Nano letters}, 18(11):6725--6730, 2018.

\bibitem{eich2018spin}
Marius Eich, Riccardo Pisoni, Hiske Overweg, Annika Kurzmann, Yongjin Lee,
  Peter Rickhaus, Thomas Ihn, Klaus Ensslin, Franti{\v{s}}ek Herman, Manfred
  Sigrist, Kenji Watanabe, and Takashi Taniguchi.
\newblock Spin and valley states in gate-defined bilayer graphene quantum dots.
\newblock {\em Physical Review X}, 8(3):031023, 2018.

\bibitem{eich2018coupled}
Marius Eich, Riccardo Pisoni, Alessia Pally, Hiske Overweg, Annika Kurzmann,
  Yongjin Lee, Peter Rickhaus, Kenji Watanabe, Takashi Taniguchi, Klaus
  Ensslin, and Thomas Ihn.
\newblock Coupled quantum dots in bilayer graphene.
\newblock {\em Nano letters}, 18(8):5042--5048, 2018.

\bibitem{overweg2017electrostatically}
Hiske Overweg, Hannah Eggimann, Xi~Chen, Sergey Slizovskiy, Marius Eich,
  Riccardo Pisoni, Yongjin Lee, Peter Rickhaus, Kenji Watanabe, Takashi
  Taniguchi, Vladimir Falko, Thomas Ihn, and Klaus Ensslin.
\newblock Electrostatically induced quantum point contacts in bilayer graphene.
\newblock {\em Nano letters}, 18(1):553--559, 2017.

\bibitem{ferrari2015science}
Andrea~C Ferrari, Francesco Bonaccorso, Vladimir FalKo, Konstantin~S Novoselov,
  Stephan Roche, Peter B{\o}ggild, Stefano Borini, Frank~HL Koppens, Vincenzo
  Palermo, Nicola Pugno, , A.~Garrido, Roman Sordan, Alberto Bianco, Laura
  Ballerini, Maurizio Prato, Elefterios Lidorikis, Jani Kivioja, Claudio
  Marinelli, Tapani Ryhanen, Alberto Morpurgo, Jonathan~N. Coleman, Valeria
  Nicolosi, Luigi Colombo, Albert Fert, Mar Garcia-Hernandez, Adrian Bachtold,
  Gregory~F. Schneider, Francisco Guinea, Cees Dekker, Matteo Barbone, Zhipei
  Sun, Costas Galiotis, Alexander~N. Grigorenko, Gerasimos Konstantatos, Andras
  Kis, Mikhail Katsnelson, Lieven Vandersypen, Annick Loiseau, Vittorio
  Morandi, Daniel Neumaier, Emanuele Treossi, Vittorio Pellegrini, Marco
  Polini, Alessandro Tredicucci, Gareth~M. Williams, Byung~Hee Hong, Jong-Hyun
  Ahn, Jong~Min Kim, Herbert Zirath, Bart~J. van Wees, Herre van~der Zant,
  Luigi Occhipinti, Andrea~Di Matteo, Ian~A. Kinloch, Thomas Seyller, Etienne
  Quesnel, Xinliang Feng, Ken Teo, Nalin Rupesinghe, Pertti Hakonen, Simon
  R.~T. Neil, Quentin Tannock, Tomas L{\o}fwander, and Jari Kinaret.
\newblock Science and technology roadmap for graphene, related two-dimensional
  crystals, and hybrid systems.
\newblock {\em Nanoscale}, 7(11):4598--4810, 2015.

\bibitem{jariwala2014emerging}
Deep Jariwala, Vinod~K Sangwan, Lincoln~J Lauhon, Tobin~J Marks, and Mark~C
  Hersam.
\newblock Emerging device applications for semiconducting two-dimensional
  transition metal dichalcogenides.
\newblock {\em ACS nano}, 8(2):1102--1120, 2014.

\bibitem{fiori2014electronics}
Gianluca Fiori, Francesco Bonaccorso, Giuseppe Iannaccone, Tom{\'a}s Palacios,
  Daniel Neumaier, Alan Seabaugh, Sanjay~K Banerjee, and Luigi Colombo.
\newblock Electronics based on two-dimensional materials.
\newblock {\em Nature nanotechnology}, 9(10):768, 2014.

\bibitem{liu2016van}
Yuan Liu, Nathan~O Weiss, Xidong Duan, Hung-Chieh Cheng, Yu~Huang, and
  Xiangfeng Duan.
\newblock Van der waals heterostructures and devices.
\newblock {\em Nature Reviews Materials}, 1(9):16042, 2016.

\bibitem{Rumelhart1988}
David~E Rumelhart, Geoffrey~E Hinton, and Ronald~J Williams.
\newblock Learning representations by back-propagating errors.
\newblock {\em Cognitive modeling}, 5(3):1, 1988.

\bibitem{Geron2017}
Aur{\'e}lien G{\'e}ron.
\newblock {\em Hands-on machine learning with Scikit-Learn and TensorFlow:
  concepts, tools, and techniques to build intelligent systems}.
\newblock " O'Reilly Media, Inc.", 2017.

\end{thebibliography}

\end{document}